# Machine Translation, Sentiment Analysis, Text Similarity, Topic Modelling, and Tweets: Understanding Social Media Usage Among Police and Gendarmerie Organizations

Emre Cihan ATES[a][1], Erkan BOSTANCI[b], Mehmet Serdar GÜZEL[c]

[a] Department of Security Science, Gendarmerie and Coast Guard Academy (JSGA), Ankara, Turkey.
[b] Department of Computer Engineering, Ankara University, Ankara, Turkey.
[c] Department of Computer Engineering, Ankara University, Ankara, Turkey.

## Abstract

It is well known that social media has revolutionized communication. Nowadays, citizens, companies, and public institutions actively use social media in order to express themselves better to the population they address. This active use is also carried out by the gendarmerie and police organizations to communicate with the public with the purpose of improving social relations. However, it has been seen that the posts by the gendarmerie and police organizations did not attract much attention from their target audience from time to time, and it has been discovered that there was not enough research in the literature on this issue. In this study, it was aimed to investigate the use of social media by the gendarmerie and police organizations operating in Turkey (Jandarma - Polis), Italy (Carabinieri - Polizia), France (Gendarmerie - Police) and Spain (Guardia Civil - Policía), and the extent to which they can be effective on the followers, by comparatively examining their activity on twitter. According to the obtained results, it was found that Jandarma (Turkey) has the highest power of influence in the twitter sample, and the findings were comparatively presented in the study.

## 1. Introduction

In recent years, social media has become an important communication platform for receiving and giving information on various topics. It is more popular than traditional communication tools (such as newspaper, television, radio) (Simeone and Russo, 2017). The most important thing that makes social media more popular than traditional means of communication is that individuals or groups can freely share their ideas and connect with others online (Ahmed et al., 2019).

---

[1]Corresponding author.
E-mail addresses: emre_cihan_ates@hotmail.com(E.C.Ates), ebostanci@ankara.edu.tr(E.Bostanci), mguzel@ankara.edu.tr(M.S.Guzel).

On the social media platform, users publish a series of messages that will introduce them to their followers. These messages are liked by the followers and can be reposted. When they are liked, they are recommended to other users who have similar inclinations through various algorithms, and if they are reposted, they can be seen by other users as well, and they spread rapidly within the social network.

Due to the rapid spread of information contained in the messages shared on social media, various individuals, groups, or organizations are trying to use social media more actively every day (Tajudeen et al., 2018). Gendarmerie and police organizations are among the institutions that use social media actively. Police organizations are structures that are responsible for protecting the public order and citizens' lives, property, and most importantly, fundamental rights and freedom. Gendarmerie organization, on the other hand, is a police force that can be a part of the armed forces in some special cases and perform similar duties to police organizations.

Social media analysis is the easiest way for individuals, institutions, or companies to understand their target audience. All over the world, gendarmerie and police organizations are increasingly using social media to learn from and engage with the public, from the perspective of improving community relations. Even though gendarmerie and police organizations that operate on a national scale actively use social media, it has been seen that their posts sometimes do not attract much attention from the public like the social media accounts of various other public institutions. The basis of the ability to communicate, which is the main purpose of social media, is to attract the attention of other people. This is because messages that attract attention, that are described as trends, are read, liked, commented on, or reposted by the users.

It is essential that the messages shared by the gendarmerie and police organizations, which are responsible for ensuring the safety of the public, attract the attention of the citizens and circulate within social media, creating an environment where communication with the public is possible. Because the widespread use of social media constitutes a potential resource in identifying problems in police-society or gendarmerie-society relations. In this study, it was aimed to investigate the use of social media by the gendarmerie and police organizations operating in Turkey (Jandarma - Polis), Italy (Carabinieri - Polizia), France (Gendarmerie - Police) and Spain (Guardia Civil - Policía), and the extent to which they can be effective on the followers, by comparatively examining their activity on Twitter. The impact was calculated through the likes and retweets of the posts, all posts were translated into English, and sentiment analysis, topic modeling, and similarity analysis were carried out. After the introduction in the study, spread of information in social media, community policing and social media, machine translation, sentiment analysis, text similarity, and topic modeling concepts were defined respectively, and the background and related work were carried out in which studies in the literature were examined.

In the methodology section, information on methods of data acquisition, pre-processing, machine translation, sentiment analysis, topic modeling, and similarity analysis were provided. In the Result and Discussion section, the results obtained after the analyses were shared and compared with other studies in the literature. In the Conclusion and Future Work section, the gains obtained after the study, some important findings and limitations, and the planned future work were presented.

## 2. Background and related work

### 2.1 Spread of information on social media

Three basic factors, sender, receiver, and medium, have an effect on the spread of information (Al-Taie and Kadry, 2017). If we think about it in terms of social media, the person who shares the post is the sender, the persons or groups expected to be affected by the post are the receivers, and the social media platform is the medium. It is known that people all over the world who use the Internet use social media for an average of 2 hours and 24 minutes per day, and this reveals how intense the online interactions are occurring in social media environment (WeAreSocial, 2020). This intense interaction is the main factor that triggers the rapid spread of information. In particular, re-sharing a message (repost, retweet) shared by a person or institution constitutes the main process in which information is spread on social media. For this reason, in the Twitter sample, the likes, comments, and reposts that a message has are the basic representations that show the power of the message in question.

### 2.2 Community Policing and Social Media

Community policing is a public security service where law enforcement agencies constantly communicate with citizens, identify the problems of citizens, and focus on solving the identified problems (Rukus et al., 2018; Williams et al., 2018). In general, the goal is communicating with citizens and preventing crime from occurring, rather than intervening after the incident occurs. In classical approaches, the police generally serve in regions based on geographic location (Somerville, 2009). However, with the increase in social media and internet usage, we are living in a time where geographical borders are removed. Especially in recent years, almost all gendarmerie and police organizations have social media accounts. Williams et al. (2018), Meijer and Thaens (2013) examined America, Meijer and Torenvlied (2016) examined the Netherlands, Fernandez et al. (2017) examined United Kingdom, and Tang et al. (2019) examined China in this context, in order to investigate the use of social media by law enforcement officers. Although the power of social media in the spread of information is known, it was seen that very few of the posts were open to two-way interaction (Bonsón et al., 2015; Hong and Nadler, 2012; Strauß et al., 2015). According to the results obtained from the studies, it was found that the gendarmerie and police organizations understood the potential of social media but did not use social media effectively enough to communicate with the public.

**2.3 Machine translation**

Machine translation, one of the most used applications of natural language processing, is a machine-driven translation of the natural language. The translation process is not an exact substitution of every word. It aims to make syntactically, grammatically and semantically correct translations between any two languages. For this reason, the translator has to analyse all the elements in the text and know how each word affects other words. Machine translations are still in development phase with apps like Google Translate, Yandex Translate, and Bing Translator, the most popular one being Google Translate (Sarkar, 2019). Aiken (2019) reviewed the translation accuracy of 50 different languages over Google translate, and as a result of the research, achieved 75% accuracy in Turkish - English translation, 80% in Spanish - English, 88% in French - English, and 90% in Italian - English. Successful results were also obtained in the studies conducted by Roca (2020), Rana and Atique (2019) and (Shekhawat, 2019) on machine translation.

**2.4 Sentiment Analysis**

Sentiment analysis is used to analyse texts and to predict feelings, thoughts, and opinions in the content of these texts. The concept of sentiment analysis is also known as idea mining, and its main purpose is to analyse people's emotions (reactions) to a particular entity (Sarkar, 2019). Within the scope of the analysis, natural language processing, statistics, machine learning, and linguistics techniques are used (Bakshi et al., 2016). Sentiments are classified as subjective or objective, positive, negative or neutral. Neutral emotions are typically at pole 0 because they do not express a particular emotion, positive emotions have a > 0 pole, and negative emotions have <0. It is possible to analyse emotion using two methods: dictionary-based approach and machine learning approach (Medhat et al., 2014). With the dictionary-based approach, a classification is made based on a sensitivity dictionary that defines the emotions expressed in text and the emotion poles of the words. In order to make analyses using the machine learning approach, training data is needed, and the information obtained as a result of the learning is verified on test data (Sarkar, 2019).

When some sentiment analysis studies in the literature that were specifically carried out on Twitter were examined, it was seen that studies on many areas were carried out such as diabetes disease (Gabarron et al., 2019), airline customer opinions (Prabhakar et al., 2019), politicians' election forecast (Huang, 2017), Brexit referendum results (Grčar et al., 2017; Shekhawat, 2019), COVID-19 (Khan et al., 2020), World Cup football tournament (Patel and Passi, 2020), Indonesia Police (Kurniawati et al., 2019), Syrian refugees (Öztürk and Ayvaz, 2018), Trump's Tweets (Sahu et al., 2020), and 5G technology (Seçkin and Kilimci, 2020). It has been determined that most of the studies used dictionary-based sensitivity analysis approach.

**2.5 Text Similarity**

The purpose of text similarity is to analyze how close text elements are to each other. In general, it is possible to perform text-similarity in two ways, lexical and semantic (Sarkar, 2019). Lexical similarity involves observing the content of the text in terms of syntax, structure, and content, and examining their similarity based on these parameters. In lexical similarity, the text becomes vectorial by performing term and document similarity. In semantic similarity, on the other hand, the goal is to identify the meanings in the text and then examine how close they are to each other. In particular, features such as POS tagging (Part-of-Speech Tagging), named entity recognition (NER), and parsing within the grammatical structure specific to each language are the tools of semantic similarity (Lane et al., 2019). Hamming, manhattan, euclidean, levenshtein arrangement, and cosine distance can be used to calculate and measure both types of similarity (Sarkar, 2019).

There are text similarity studies in the literature such as online article evaluation (Lahitani et al., 2016), court case documents (Novotná, 2020), legal documents (Thenmozhi et al., 2017), news (Singh and Singh, 2020), comments and criticisms on Twitter (Prasetyo and Winarko, 2017), article similarity (Dunn et al., 2018), emergency detection via social media (Huang et al., 2020). While the main subject that should be evaluated in text-similarity should be semantic, it is not very easy to make these analyses as a result of the almost infinite variations of usage in each language. In lexical methods, similar words such as "book" and "cook" make the analysis difficult, and it is considered that different methods such as weighting of words can be used to overcome this. It has been observed that the cosine distance method is used in the similarity calculation in most of the studies on text similarity.

**2.7 Topic modeling**

Topic modeling involves the use of mathematical and statistical modeling techniques to extract key issues and themes from a document corpus. Topic modeling, also commonly known as probabilistic statistical models, uses a variety of techniques such as Latent Semantic Indexing (LSI), Singular Valued Decomposition (SVD), Latent Dirichlet Allocation (LDA), and probabilistic latent semantic indexing (PLSI) to discover connected hidden semantic structures in text data (Lane et al., 2019). The most popular of these techniques is LDA, and it is a topic modeling algorithm that recognizes hidden word formation patterns through uncontrolled word distribution in documents (Jacobi et al., 2016; Zhou and Zhang, 2017). Its operation is within the framework of the Bayesian approach, and the subject mix is the Dirichlet distribution (Jelodar et al., 2019). Therefore, the document consists of a finite mix of topics, and each subject has a possibility for document representation. It is possible to visualize the subject models interactively with pyLDAvis, and they can be visualized in two dimensions with size reduction techniques such as t-SNE, PDA, and MDS (Sarkar, 2019).

When some text similarity studies on topic modeling in the literature are examined, it is seen that there are studies such as identifying the topics on Twitter and Reddit (Curiskis et al., 2020), COVID-19 topic modeling with Twitter (Kaila and Prasad, 2020), identifying suggestion and reference topics (Basuki et al., 2019), subject analysis from political sentences on Twitter (Sarddar et al., 2020), identifying the topics in journalistic texts (Jacobi et al., 2016), and topic modeling of comments on Disneyland (Luo et al., 2020).

## 3. Methodology

In this study, the goal was to examine the social media activity of gendarmerie and police organizations in Turkey, Italy, France, and Spain on Twitter and to comparatively examine the messages they give on social media using text mining methods on a common platform.

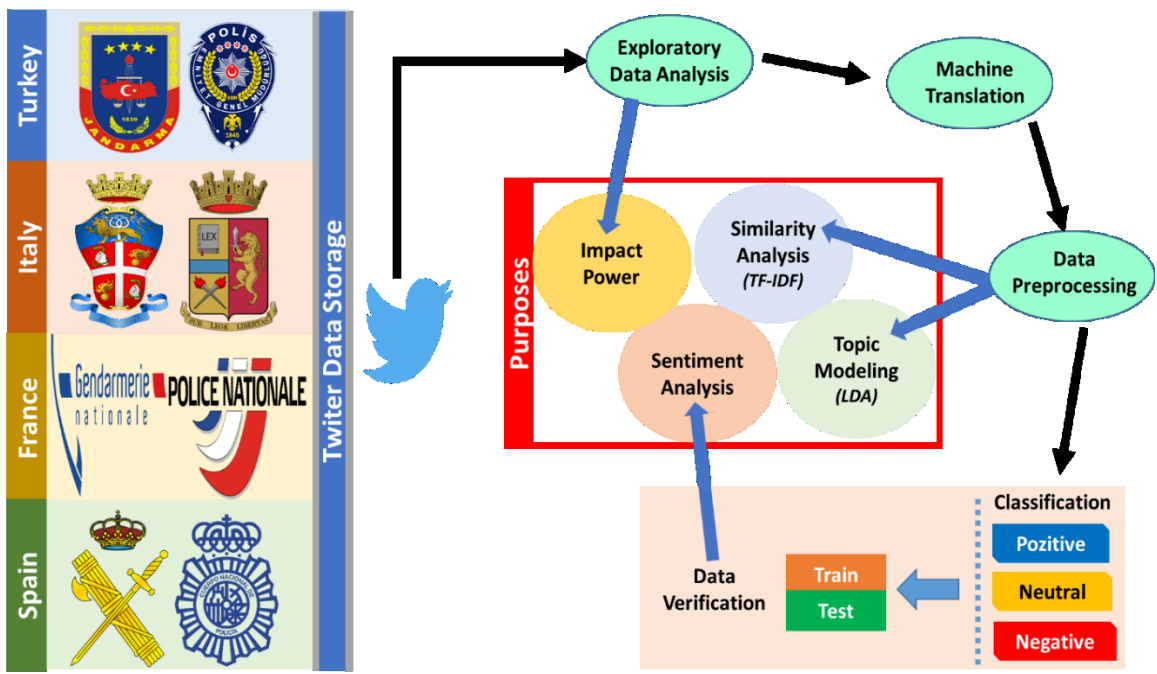

**Fig. 1.** Analysis model for Gendarmerie and Police organizations

This study was conducted with the model in fig. 1, and the procedures regarding the methodology are explained in the subheadings.

### 3.1. Data Collection

The data used in this study consisted of the Tweets of the gendarmerie and police organizations in Turkey (TR), Italy (IT), France (FR), and Spain (ES). A Twitter Developer account was opened and Twitter Rest API was used in order to obtain the social media data of the gendarmerie and police organizations at the international level.

**Table 1.** Data retrieved from Twitter

| Country | Name of Organization | Organization Type | Username | Tweets | Start Date | End Date | Language |
|---|---|---|---|---|---|---|---|
| Turkey | Jandarma | Gendarmerie | @jandarma | 3071 | 02.03.2017 | 08.03.2020 | Turkish |
| Turkey | Polis | Police | @EmniyetGM | 3228 | 18.01.2018 | 08.03.2020 | Turkish |
| Italy | Arma dei Carabinieri | Gendarmerie | @_Carabinieri_ | 3232 | 21.12.2017 | 09.03.2020 | Italian |
| Italy | Polizia di Stato | Police | @poliziadistato | 3246 | 26.03.2018 | 08.03.2020 | Italian |
| France | Gendarmerie Nationale | Gendarmerie | @Gendarmerie | 3239 | 21.10.2017 | 09.03.2020 | French |
| France | Police Nationale | Police | @PoliceNationale | 3141 | 21.08.2018 | 08.03.2020 | French |
| Spain | Guardia Civil | Gendarmerie | @guardiacivil | 3243 | 16.09.2019 | 08.03.2020 | Spanish |
| Spain | Policía Nacional | Police | @Policía | 3227 | 26.05.2019 | 09.03.2020 | Spanish |

Tweets in Turkish, Italian, French, and Spanish were collected from the Gendarmerie and Police departments as shown in Table 1. When the usernames were examined, it was seen that the accounts of all institutions were real and had a blue tick, in other words, they were verified by Twitter. The maximum number of tweets can be around 3,200, and within the scope of the restrictions imposed by Twitter, the number of tweets (max_count) that can be taken for each data retrieval request can be 200 at most. In the data obtained, there is the LIFO (Last in First Out) principle, and the dates of the tweets obtained due to the maximum possible tweet restriction are shown in Table 1. The tweets subject to the analysis were retrieved on March 9, 2020, and the differences in the dates are related to how often the institutions use Twitter. Since the data were in four different languages, UTF-8 was used in character encoding. The data were taken from Twitter in JSON format and recorded in CSV format.

### 3.2. Exploratory Data Analysis (EDA) and Machine Translation

On the data collected from the Twitter accounts of the Gendarmerie and Police organizations, the frequency of tweeting and the follower ratios proportional to the population of the country (for data normalization purposes) were calculated. Then, like (total number of favourites/number of followers) and retweet (total number of retweets/number of followers) impact was calculated. In order to make a more detailed comparison on the results obtained after exploratory data analysis, 25.627 tweets (6.299 tweets in Turkish, 6.478 tweets in Italian, 6.380 tweets in French, and 6.470 tweets in Spanish) in four different languages were translated into English by machine translation. In translation into English language, TextBlob API was used. The translation cycle was established with TextBlob, and the information in translation was received via Google Translate API.

### 3.3. Data Preprocessing

In data preprocessing, it was aimed to transform the texts in the data set into a common form. For these reasons, tweets were converted to lowercase letters and unnecessary content (URLs, space pattern, numbers and special characters, Twitter mentions, punctuation, retweet symbols, and stopwords) was removed from text. NLTK API was used in order to remove stopwords. Lemmatization was carried out using WordNetLemmatizer on NLTK before topic modeling.

### 3.4. Sentiment Analysis and machine learning

Sentiment analysis is used to identify feelings or ideas in unstructured large volumes of data. Sentiment analysis includes three types of polarity: negative, neutral, and positive. In this study, word-based approach was used, and the polarity of each tweet was determined by assigning a score from -1 to 1 according to the words used in it. A negative score means a negative emotion, and a positive score means a positive emotion. There is also a subjectivity score assigned to each tweet. The subjectivity score is between 0 and 1, and a value close to 0 means objective and a value close to 1 means subjective. In this study, sentiment and subjectivity analyses were carried out on TextBlob API using unigrams. In order to evaluate the performance of the sentiment analysis obtained, the data belonging to each organization were divided into 75% training and 25% test according to their sentiment status. Each tweet in the training data set was tested on test data by trying to learn it with the random forest classification algorithm. F1 Score, Accuracy, Precision, and Recall values were used for evaluation.

### 3.5. Similarity Analysis and Topic Modeling

The lexical method was preferred for the similarity analysis between the gendarmerie and police organizations. TF-IDF (Term Frequency - Inverse Document Frequency) method, in which the weight of each word is calculated, was used to access information through text data. After the documents to be analysed were made vectorial, the cosine similarity was used in the vector space model.

Unsupervised learning was applied to find the word groups in the data called "topics" in the tweets. Latent Dirichlet Allocation (LDA) was used in the modeling (max_iter = 5, n_components = 10 top_n = 10). pyLDAvis was used for the display of the detected topics at the vectorial level.

### 3.7. Experimental Settings

In this study, all operations were carried out in Python 3.8 using a Windows PC powered with 12 GB RAM. All modeling and algorithms were programmed in Python, using Pycharm. Tweepy, NLTK, Numpy, Pandas, TextBlob, Sklearn, WordNetLemmatizer, Seaborn and Matplotlib libraries were used in the acquisition, preprocessing, and modeling of the tweets.

### 4. Result

Exploratory data analysis was conducted in order to identify the social media usage habits of the gendarmerie and police organizations operating in Turkey, Italy, France, and Spain, and to identify the effect of their posts on social media. In the exploratory data analysis, the data belonging to different countries and law enforcement agencies were normalized with the population of the country and the number of followers for comparison.

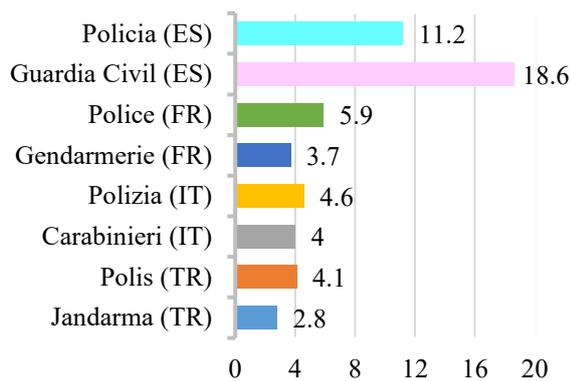

**Fig. 2.** Tweeting frequency (Daily average)

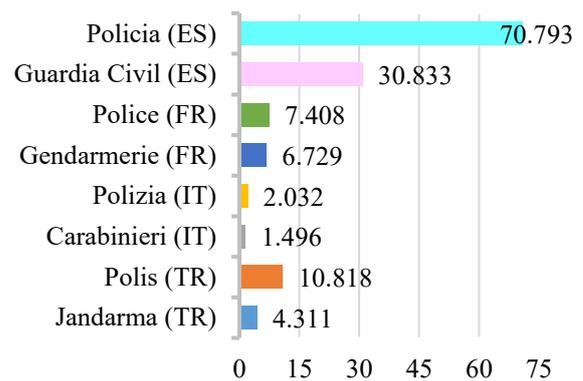

**Fig. 3.** Number of followers (Proportional to country population*1000)

As can be seen in fig. 2 and fig. 3, as a result of the exploratory data analysis performed on the data of the gendarmerie and police organizations, it was found that;

- The average number of tweeting was especially high in Policía (ES) and Guardia Civil (ES), and all other law enforcement agencies posted below average,
- The least tweeting with an average of 2.8 tweets per day was in Jandarma (TR),
- Law enforcement agencies posted an average of 6.8 tweets per day,
- Policía (ES) had the highest number of followers in proportion to the population of the country, and the Guardia Civil (ES) also had a significant number of followers in the same country,
- Carabinieri (IT) had the least number of followers proportional to the population of the country, and the number of followers of Polizia (IT), in the same country, was lower than other countries.

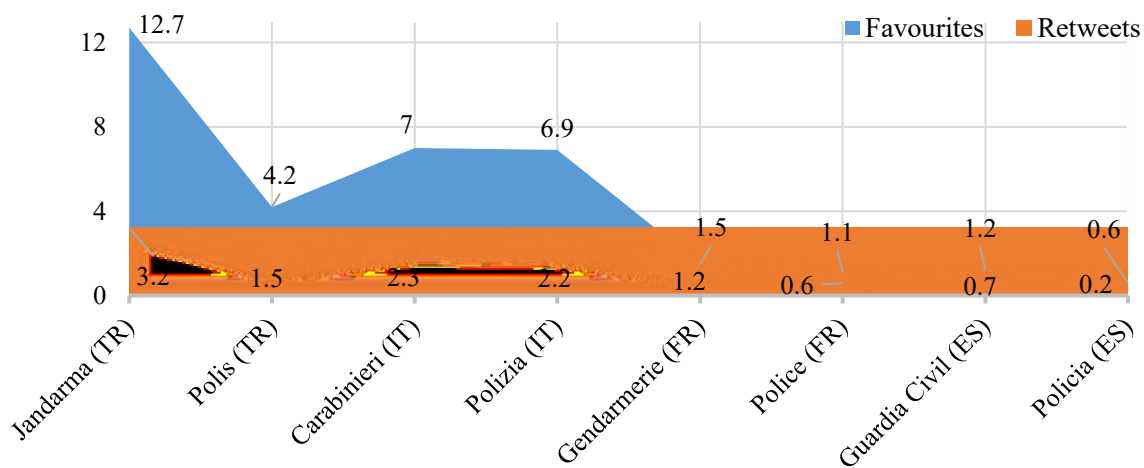

**Fig. 4.** The impact of tweets (In proportion to the number of followers)

The total number of likes and retweets of the posts in the Twitter accounts of the gendarmerie and police organizations were proportioned to the number of followers, and the power of impact (favourites/retweets) was calculated as shown in Figure 4. In this context, as shown in Figure 4, it was found that Jandarma (TR) had the highest rate of avourites and retweets, while Policía (ES) had the lowest rate of favourites and retweets.

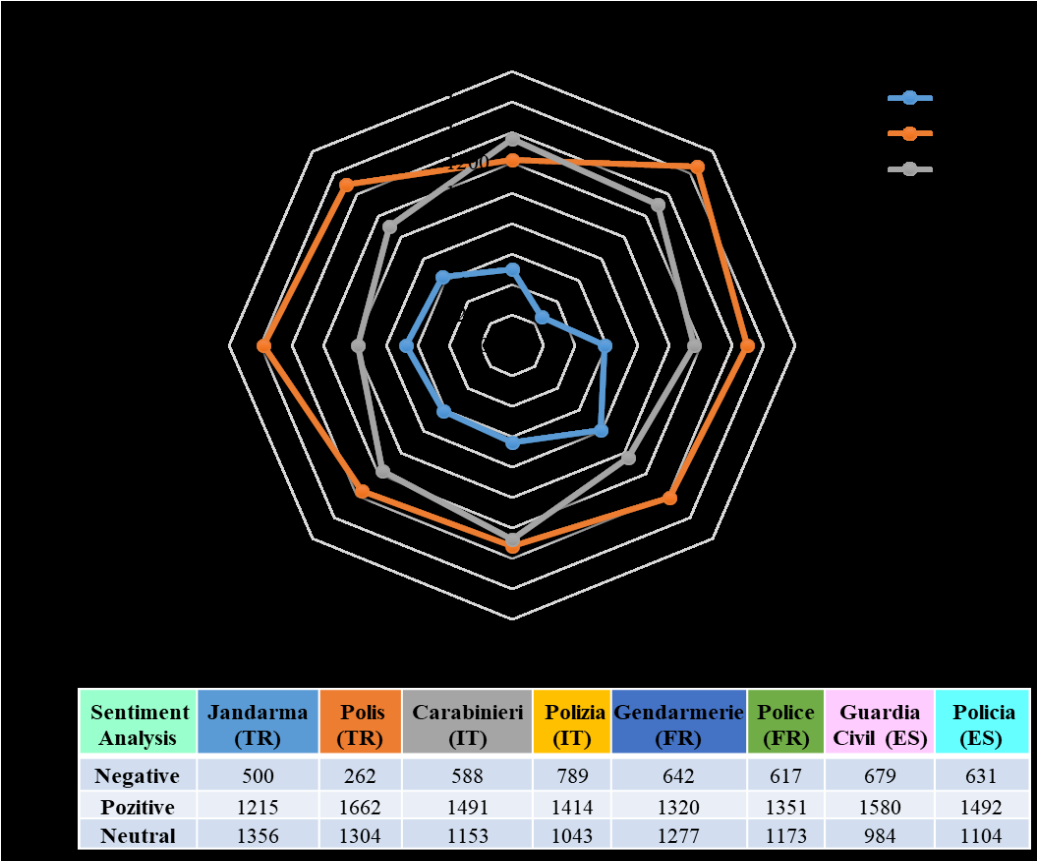

**Fig. 5.** Sentimental analysis of the tweets

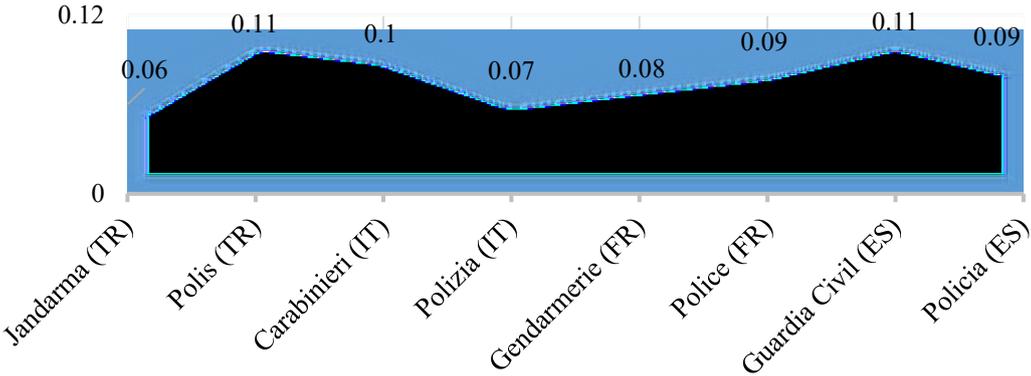

**Fig. 6.** Polarity values of the tweets

The values obtained as a result of sentimental analysis of the tweets belonging to the gendarmerie and police organizations are shown in fig 5. In fig 6, polarity values were averaged. When the data in fig. 5 and fig. 6 are examined, it is seen that;

- The organization with the highest average polarity value were Polis (TR) and Guardia Civil (ES),

- The organization with the lowest average polarity value was the Jandarma (TR),
- Carabinieri (IT) shared more positive messages compared to Polizia (IT) in Italy,
- In France, Police (FR) shared more positive messages compared to Gendarmerie (FR),
- In Spain, the Guardia Civil (ES) shared more positive messages compared to Policía (ES).

**Table 2.** Performance values of random forest classification modeling

| Models | Train | Test | F1 Score | Accuracy | Precision | Recall |
| --- | --- | --- | --- | --- | --- | --- |
| Jandarma (TR) | | | 82.724 | 86.979 | 81.074 | 85.817 |
| Polis (TR) | | | 79.820 | 89.591 | 77.970 | 83.675 |
| Carabinieri (IT) | | | 75.349 | 79.208 | 74.863 | 78.810 |
| Polizia (IT) | 75 % | 25 % | 78.456 | 79.680 | 78.616 | 80.623 |
| Gendarmerie (FR) | | | 74.719 | 77.407 | 73.507 | 79.538 |
| Police (FR) | | | 80.152 | 81.170 | 78.864 | 84.547 |
| Guardia Civil (ES) | | | 80.937 | 82.614 | 80.959 | 82.870 |
| Policía (ES) | | | 84.366 | 86.245 | 83.812 | 86.698 |
| Average | | | 79.565 | 82.862 | 78.708 | 82.822 |

The random forest classification algorithm was used to identify the machine learning validity of the polarity analysis of the gendarmerie and police organizations. F1 Score, accuracy, precision, and recall values in table 2 were used for the evaluation. When the values obtained were examined, F1 score was found that despite machine translation, the polarity analysis revealed compatible results between 74% and 84%.

Correlation analysis was carried out in order to examine the effect of the sentiment status of the posts on retweets and favourites, and the results obtained are presented in fig. 7. When the data in fig. 7 are examined, it is seen that;

- While there is a negative relationship in Polis (TR) (- 0.021), Jandarma (TR) (- 0.014), and Polizia (IT) (- 0.0081) in the favourites of positive posts, the relationship is positive in other gendarmerie and police organizations (The highest relationship Police (FR) \ (0.077)),
- While there is a negative relationship in Police (FR) (- 0.025) and Carabinieri (IT) (- 0.0096) in the favourites of negative posts, the relationship is positive in other gendarmerie and police organizations (The highest relationship is Police (FR) \ (0.063)),
- While there is a positive relationship in Polis (TR) (0.0082) for the favourites of the posts with neutral content, the relationship is negative in the other gendarmerie and police organizations (The lowest relationship in Carabinieri (IT) \ (- 0.063)),
- While there is a positive relationship in Police (FR) (0.0098) and Policía (ES) (0.001) in the retweeting of positive content, the relationship is negative in other gendarmerie and police organizations (The lowest relationship is Polis (TR) (- 0.077)),

- While there is a negative relationship in Carabinieri (IT) (- 0.0096), Police (FR) (- 0.0061), and Guardia Civil (ES) (- 0.044) in retweeting of negative content, the relationship is positive in other gendarmerie and police organizations (Highest relationship in Polizia (IT) \ (0.044)),

- While there is a negative relationship in Gendarmerie (FR) (- 0.011), Police (FR) (- 0.0051), and Policía (ES) (- 0.034) in retweeting of neutral content, the relationship is positive in other gendarmerie and police organizations (Highest relationship in Polis (TR) and Guardia Civil (ES) \ (0.063)),

- The average rate of retweeting a liked post was 0.55 in the gendarmerie and police departments, the highest relation was found in Policía (ES) (0.83), and the lowest relation was in the Guardia Civil (ES) (0.31).

It was found that different emotional states had different effects on retweets and likes in the gendarmerie and police departments; some had positive interactions, while others had negative interactions. No direct effect of the frequency of posting and the number of followers was observed on impact (Fig. 1, 2, 3). For this reason, it was attempted to understand how similar the posts were in order to examine the subject in more detail using text mining methods.

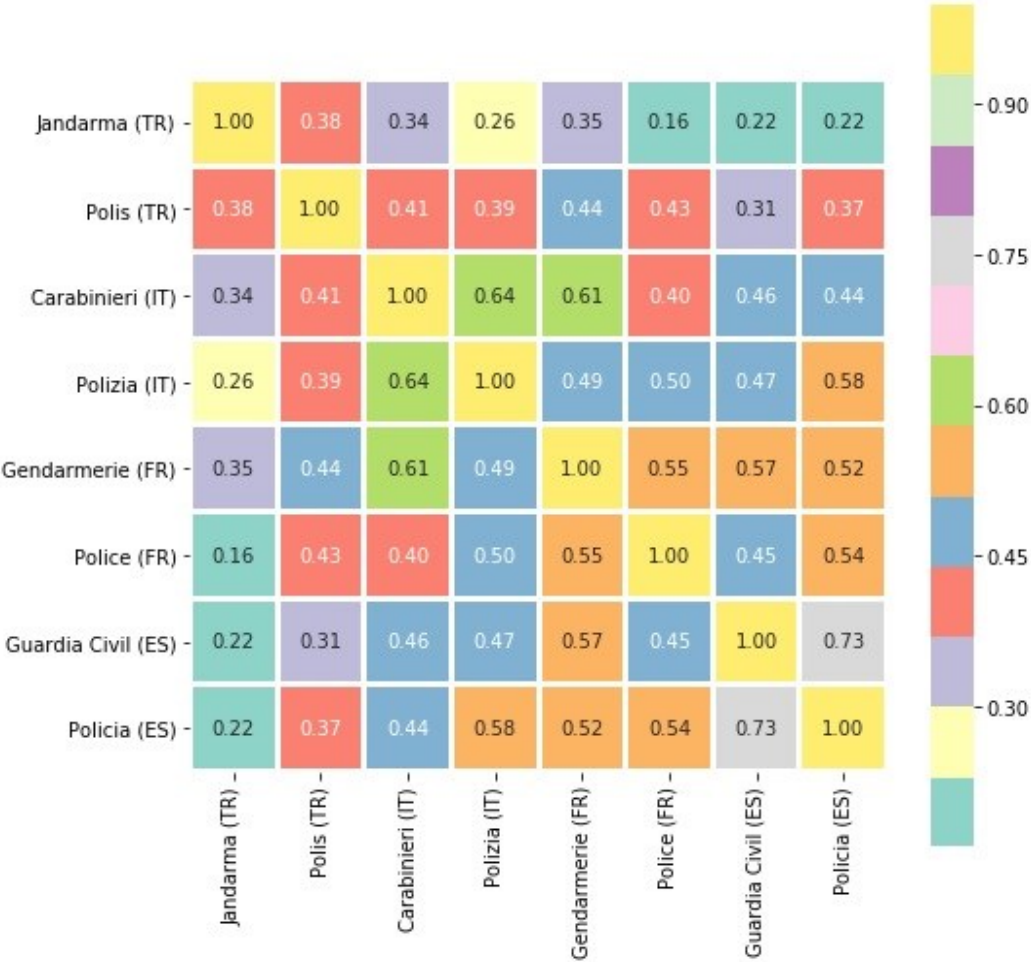

**Fig. 8.** Posting similarity matrix (TF-IDF)

Similarity was calculated using the TD-IDF method after lemmatization was performed on the text translated into English. The information obtained is presented on the matrix in fig. 8. When the content similarity analysis presented in fig.8 is examined, it is seen that;

- Jandarma (TR)'s contents were similar to those of Polis (TR) by 38%,
- Polis (TR)'s contents were similar those of Gendermarie (FR) by 44%,
- Carabinieri (IT)'s contents were similar to those of Polizia (IT) by 64%,
- Gendarmerie (FR)'s contents were similar to those of Carabinieri (IT) by 61%,
- Police (FR)'s contents were similar to those of Gendarmerie (FR) by 55%,
- Guardia Civil (ES)'s contents were similar to those of Policía (ES) by 74%,
- According to the content similarity ratio, the most original posts were from Jandarma (TR) (between 0.16 - 0.38),
- When the rate of similarity of the gendarmerie and police organizations within the countries were examined, the lowest similarity was found in Turkey (TR) (0.38), and it was seen that France (FR) (0.55), Spain (ES) (0.73) and Italy (IT) (0.64) had higher similarity rates.

In order to better understand the use of social media by the gendarmerie and police organizations, it was thought that the subjects they shared should be revealed. Machine learning-based LDA (Latent Dirichlet Allocation) analysis and distance mapping between subjects were carried out for subject analysis. The results obtained from the LDA analysis are given in Table 23. When the data in question are examined, it can be said that Gendarmerie (FR) and Carabinieri (IT) used the word "Military," Jandarma (TR) used the word "Terrorist," and Polizia (IT) used the word "Mafia." It is thought that the data in Table 3 varied according to the cultural, social, and economic structure of the society in which the gendarmerie and police organization are located and the management style of the institution.

## 5. Discussion

Twitter is more content-centered than relation-centered, which distinguishes it from other social media applications (Asadi and Agah, 2018). At the same time, the lack of indicators showing reciprocal relations in the application makes it difficult to measure the impact of tweets. However, in the studies conducted, the number of user followers, analysis of the content of the tweets, favourites, retweets, and comments are some of the basic elements that show that the user has an impact (Dahka et al., 2020; Garcia et al., 2017; Rezaie et al., 2020). In this study, all parameters except comments were examined with text mining techniques. There are two important reasons for not conducting reviews on the comments; the first is the difficulty arising from the text containing Turkish, French, Italian, and Spanish content, and the second is the difficulty of machine translation of the social media language, where abbreviated expressions are widely used by the users.

Gendarmerie and police organizations convey their feelings and thoughts on various issues to their followers, usually on the basis of crime and security. The main purpose of communication is that the messages conveyed are perceived by the target audience. Especially in Spain, although law enforcement units are quite willing to inform the public by posting more tweets, it was found that gendarmerie and police organizations in other countries are less willing to inform citizens. However, although the units in Spain tweet quite a lot on average daily, it is understood that the tweets are not effective enough on the citizens. The weak power of impact is one of the most fundamental factors affecting the speed of the spread of information on social media. The parameters between frequent tweeting and having an impact are not fully known (Riquelme and González-Cantergiani, 2016). However, in our study, it was found that Jandarma (TR), which tweeted the least with an average of 2.8 tweets per day, had the highest impact value.

In communication, it is known that negative stimuli attract more attention than positive stimuli (Kamp et al., 2015), and similar results were obtained in the studies conducted by Trevors and Kendeou (2018) and Kätsyri et al. (2016). In this study, it was found that the posts with low polarity average, that is, the posts containing more negative content compared to other organizations, had a higher impact on the citizens. It was found that messages with negative content, other than Police (FR), generally created favourites or retweets in a positive correlation, and it was found that police received more interaction from positive messages. In addition, it was found that the favourite and retweet correlation in the posts on social media is very strong, especially in Policía (ES).

The authenticity of the topics shared on social media in a way that attracts the attention of the followers is one of the important parameters affecting the power of impact (Chen et al., 2020; Hamzah et al., 2020). In this study, it was found that the organization with the lowest text similarity rate was Jandarma (TR), and topic models were created. In topic modeling, it was found that social and cultural situations are effective factors in determining the topics.

## 6. Conclusion and Future Works

With the widespread use of social media today, individuals and institutions can freely express their opinions, thoughts, and even their attitudes. In this sense, the gendarmerie and police organizations also express their feelings and thoughts on social media, and it was found that their posts did not receive enough attention from the citizens from time to time. The posts on social media need to attract attention, in other words, have an impact on people, for the spread of the information. In this context, many analysis methods of data mining (Machine translation, sentiment analysis, machine learning, correlation, text similarity, topic modeling) were used in our study, and the results obtained were compared at national/international dimensions. It is considered that the study differs from many studies in the field of text mining since it is specific to the gendarmerie and police organizations working in the field of combating crime. In this study, the purpose was to investigate the use of social media of the gendarmerie and police organizations operating in Turkey (Jandarma - Polis), Italy (Carabinieri - Polizia), France (Gendarmerie - Police) and Spain (Guardia Civil - Policía), and to what extent they can be effective on the followers, by comparatively examining their activity on Twitter. As a result of the analysis carried out, it was found that Jandarma (TR) used Twitter very effectively, with the lowest positive polarity value of the tweets, the least frequent use of Twitter, and the most original content sharing compared to all other law enforcement units examined in this study. In addition, within the scope of the analyses, Turkish, Italian, French, and Spanish tweets were translated into English with machine translation, and the emotional states formed after sentimental analysis were tested with machine learning. Although it was machine translation, it is remarkable that the average of the results obtained with random forest classification achieved 79.565% F1 score.

Some limitations should be taken into account while generalizing the results of this study. The first of these limitations is that the study was conducted exclusively on Twitter and the second is that it covered a limited time period. In this context, it is aimed to examine the posts of the gendarmerie and police organizations on Facebook and Instagram pages as well as Twitter in future studies.

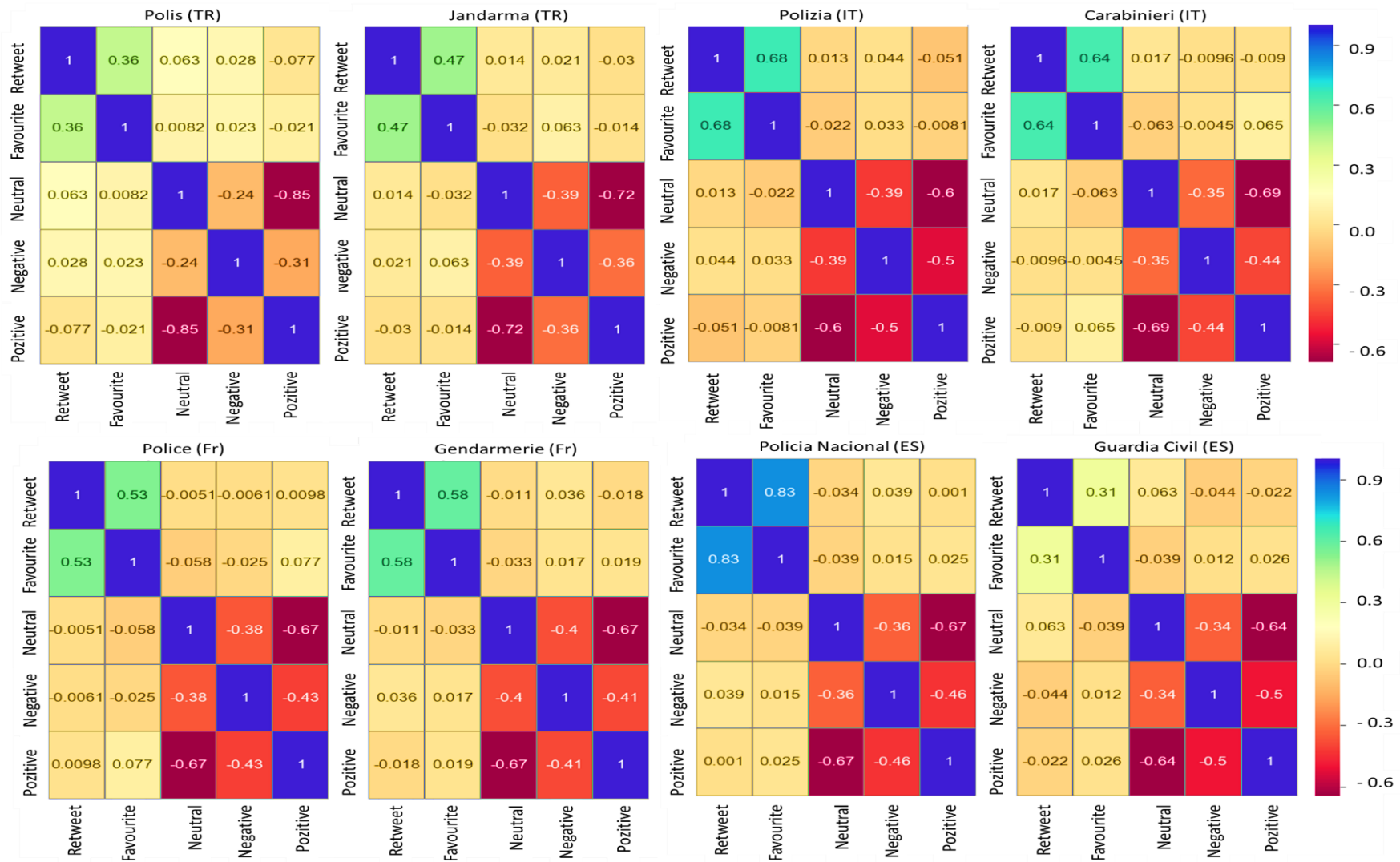

**Fig. 7.** Sentiment-based correlation of Gendarmerie and Police organizations

**Table 3.** Intertopic distance map and topic modeling

| Organizations / Intertopic Distance Map | | Topic 1 | Topic 2 | Topic 3 | Topic 4 | Topic 5 | Topic 6 | Topic 7 | Topic 8 | Topic 9 | Topic 10 |
|---|---|---|---|---|---|---|---|---|---|---|---|
| Jandarma (TR) | 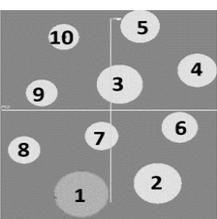 | operation, carry, terrorist, command, provincial, gendarmerie, countryside, neutralize, seize, weapon | gendarmerie, command, training, commando, school, center, complete, general, day, successfully | suspect, search, command, carry, provincial, operation, seize, gendarmerie, district, catch | suspect, drug, seize, search, person, make, vehicle, application, gendarmerie, istanbul | gendarmerie, general, visit, commander, command, provincial, minister, arif, çetin, martyr | gendarmerie, command, guard, ceremony, commander, soldier, provincial, coast, hold, province | traffic, winter, safe, country, citizen, let, helicopter, come, van, vehicle | student, school, happy, world, national, brother, championship, congratulation, woman, primary | nation, driver, love, holiday, wish, life, accident, live, time, future | anniversary, veteran, martyr, loss, sorry, hero, year, mustafa, homeland, kemal |
| Polis (TR) | 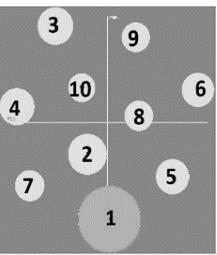 | general, police, visit, security, chief, manager, information, directorate, dear, minister | hold, minister, information, meeting, ceremony, police, special, interior, operation, coordination | traffic, accident, driver, control, speed, vehicle, safe, organization, rule, holiday | information, carry, country, detailed, inspection, application, child, simultaneously, people, personnel | operation, day, unit, drug, narcotic, fight, carry, street, determine, crime | police, year, nation, istanbul, officer, thanks, training, student, woman, center | ankara, anniversary, martyr, life, love, mustafa, veteran, kemal, respect, road | peace, citizen, use, security, duty, today, city, team, belt, agency | gendarmerie, family, child, headquarters, pedestrian, command, happy, occasion, commander, mardin | guard, world, news, program, peaceful, seize, neighborhood, june, bazaar, support |
| Carabinieri (IT) | 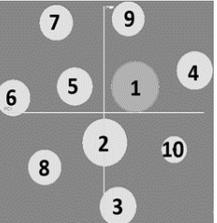 | arrest, people, drug, carry, precautionary, measure, company, association, station, seize | general, weapon, commander, good, nistri, giovanni, visit, gen, carabinieri, minister | protection, minister, heritage, cultural, force, event, police, command, collaboration, work | italian, world, carabiniere, center, team, sport, athlete, congratulation, cup, championship | car, waste, control, place, gold, medal, seize, del, choose, area | weapon, order, family, activity, regiment, police, new, mission, train, discover | day, year, thanks, wish, today, military, great, life, commitment, make | operation, food, anti, health, weapon, territory, international, agri, environmental, ndrangheta | weapon, anniversary, historical, helicopter, ceremony, open, young, live, squadron, want | condolence, express, european, road, chief, nature, come, accident, environment, action |
| Polizia (IT) | 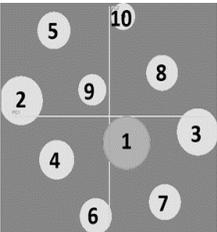 | arrest, drug, operation, criminal, investigation, association, mafia, deal, robbery, crime | year, thousand, police, italian, school, train, new, station, headquarters, identify | police, policeman, work, year, state, project, country, day, crime, meet | victim, men, life, people, accident, road, safety, year, memory, drive | italy, car, arrest, vehicle, theft, steal, good, stop, citizen, use | security, case, center, social, network, public, investigation, sign, agreement, prevention | child, traffic, road, home, agent, section, family, city, blood, trieste | medal, win, gold, man, bronze, team, silver, champion, final, podium | today, rome, concert, piazza, naples, journalist, san, celebrate, direction, gabrielli | president, travel, follow, shot, history, free, today, helicopter, presentation, image |

| Organizations / Intertopic Distance Map | | Topic 1 | Topic 2 | Topic 3 | Topic 4 | Topic 5 | Topic 6 | Topic 7 | Topic 8 | Topic 9 | Topic 10 |
|---|---|---|---|---|---|---|---|---|---|---|---|
| Gendarmerie (FR) | 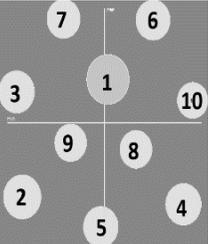 | gendarmerie, gendarme, discover, national, new, daily, year, school, intervention, mission | arrest, year, individual, theft, group, investigation, gendarme, criminal, center, state | department, general, place, lizurey, national, family, thank, come, richard, army | day, gendarme, child, today, police, night, force, student, france, control | traffic, seize, arrest, end, individual, dismantle, cannabis, area, operation, gendarme | advice, make, stay, return, leave, protect, start, minister, good, information | follow, road, security, amp, vehicle, car, death, drive, driver, comrade | colonel, morning, population, meet, mountain, use, live, summer, come, rescue | gendarme, safety, mobilize, time, respect, allow, operation, little, man, action | good, team, world, near, open, military, officer, congratulation, champion, competition |
| Police (FR) | 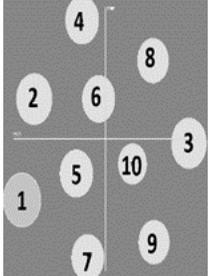 | police, tip, officer, case, prepare, rescue, scam, academy, new, day | police, arrest, join, network, meet, welcome, follow, officer, news, miss | report, content, internet, illegal, violence, platform, child, sexual, victim, police | police, day, make, emergency, know, hello, service, open, policeman, specially | video, report, thank, investigator, relay, mobilize, longer, violence, investigation, identify | hello, thank, vigilance, platform, pharos, good, station, police, evening, complaint | participate, online, use, know, dissemination, end, peace, security, advice, avoid | police, officer, year, discover, crime, help, profession, animal, work, action | site, sure, brigade, risk, great, dont, forget, make, prevention, weekend | congratulation, safety, come, road, behavior, night, event, question, answer, ensure |
| Guardia Civil (ES) | 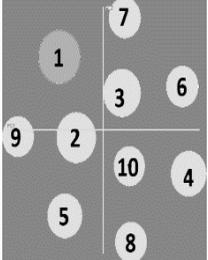 | dog, carry, know, rescue, mountain, information, work, guide, mobile, detect | guard, civil, today, challenge, life, national, receive, league, friend, cyberspace | good, locate, thanks, important, condition, collaboration, dissemination, disappearance, news, security | help, cost, lot, know, way, urgent, advice, act, road, activity | day, need, avoid, phase, sign, say, enjoy, start, road, allow | use, car, road, control, dont, vehicle, right, come, let, think | traffic, year, light, vehicle, continue, follow, drive, crime, driver, video | work, service, old, year, false, message, environment, prevent, water, play | place, child, use, cause, wait, family, rest, people, new, minor | like, look, animal, know, end, stop, leave, think, step, dog |
| Policía (ES) | 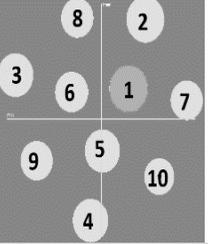 | need, remember, good, protect, time, help, leave, start, dont, lose | police, agent, security, national, year, today, work, ready, woman, officer | follow, police, criminal, avoid, tip, operation, station, want, miss, network | arrest, come, man, people, info, year, car, door, home, stop | crime, send, information, victim, ask, bank, email, child, money, fight | life, work, enjoy, respect, dont, day, case, forget, write, good | check, look, make, leave, link, sure, ignore, bank, real, appointment | want, animal, know, like, love, school, family, need, participate, care | detain, victim, detainee, woman, tell, organization, agent, dismantle, gender, sexual | day, thanks, share, tomorrow, drug, help, traffic, safety, return, face |